\documentclass[aps,prb,twocolumn,groupedaddress,amssymb,superscriptaddress]{revtex4-1}
\usepackage[pdftex]{graphicx}%_0010  % per pdflatex scommentare questa riga
\usepackage{color}
\definecolor{dg}{rgb}{0,0.5,0} % darkgreen
\usepackage{bm}      % RevTex4 bold math

\newcommand{\cefaof}{Ce\-Fe\-As\-O$_{1-x}$\-F$_{x}$}
\newcommand{\refaof}{RE\-Fe\-As\-O$_{1-x}$\-F$_{x}$}

\begin{document}

%\preprint{}

\title{Nanoscopic coexistence of magnetic and superconducting states \\ within the FeAs layers of Ce\-Fe\-As\-O$_{1-x}$\-F$_{x}$}



\author{S.~Sanna}
\email[]{Samuele.Sanna@unipv.it}
\affiliation{Dipartimento di Fisica ``A.\ Volta'' and Unit\`a CNISM di Pavia, I-27100 Pavia, Italy}
\author{R.~De Renzi}
\affiliation{Dipartimento di Fisica and Unit\`a CNISM di Parma, I-43124 Parma, Italy}
\author{T.~Shiroka}
\affiliation{Laboratorium f\"ur Festk\"orperphysik, ETH-H\"onggerberg, CH-8093 Z\"urich, Switzerland}
\affiliation{Paul Scherrer Institut, CH-5232 Villigen PSI, Switzerland}
\author{G.~Lamura}
\affiliation{CNR-SPIN and Universit\`a di Genova, via Dodecaneso 33, I-16146 Genova, Italy}
\author{G.~Prando}
\affiliation{Dipartimento di Fisica ``A.\ Volta'' and Unit\`a CNISM di Pavia, I-27100 Pavia, Italy}
\affiliation{Dipartimento di Fisica ``E.\ Amaldi'', Universit\`a di Roma3-CNISM, I-00146 Roma, Italy}
\author{P.~Carretta}
\affiliation{Dipartimento di Fisica ``A.\ Volta'' and Unit\`a CNISM di Pavia, I-27100 Pavia, Italy}
\author{M.~Putti}
\affiliation{CNR-SPIN and Universit\`a di Genova, via Dodecaneso 33, I-16146 Genova, Italy}
\author{A.~Martinelli}
\affiliation{CNR-SPIN Corso Perrone 24, I-16146 Genova, Italy}
\author{M.R.~Cimberle}
\affiliation{CNR-IMEM, via Dodecaneso 33, I-16146 Genova, Italy}
\author{M.~Tropeano}
\affiliation{CNR-SPIN and Universit\`a di Genova, via Dodecaneso 33, I-16146 Genova, Italy}
\author{A.~Palenzona}
\affiliation{CNR-SPIN and Universit\`a di Genova, via Dodecaneso 33, I-16146 Genova, Italy}

%\thanks{}


\date{\today}

\begin{abstract}
We report on the coexistence of magnetic and superconducting states in \cefaof\ for $x=0.06(2)$, characterized by transition temperatures $T_m=30$ K and $T_c=18$ K, respectively. Zero and transverse field muon-spin relaxation measurements show that below 10 K the two phases coexist within a nanoscopic scale over a large volume fraction. This result clarifies the nature of the magnetic-to-superconducting transition in the \cefaof\ phase diagram, by ruling out the presence of a quantum critical point which was suggested by earlier studies.
\end{abstract}

%\pacs{}
%\keywords{}

\maketitle

% References should be done using the \cite, \ref, and  \label commands

The recent discovery of high-$T_c$ superconductivity (SC) close to the disruption of magnetic (M) order in Fe-based compounds has stimulated the scientific community to further consider the role of magnetic excitations in the pairing mechanism.
In order to address this point it is necessary to understand how the ground state evolves from the M to the SC phase within each family of Fe-based superconductors. In the \refaof\ family (hereafter RE1111, with RE=La or a rare earth) early experiments have suggested that the M-SC crossover is RE-dependent. For instance, a smooth reduction of the magnetic and superconducting ordering temperatures, $T_m$ and $T_c$ respectively, was found for RE=Ce,\cite{Zhao2008} suggesting the presence of a quantum critical point.\cite{Amato} For RE=Sm a partial coexistence of the M and SC states was found,\cite{Drew2009} while a first order transition seems to occur for RE=La.\cite{Luetkens2009}
Successive studies \cite{Sanna2009,Hess2009,Kamihara2010} have shown that the doping region where $T_m$ and $T_c$ are both non-zero is virtually point-like in Sm1111, demonstrating that the cases of RE=La and Sm can be reconciled under a unique behavior.\cite{Sanna2009} Recently,
nanoscale electronic inhomogeneities have been shown to be present in both RE=La and Sm in a wide range above the crossover region.\cite{Lang2010} Actually also the case of RE=Ce is susceptible to further investigation concerning the presence of electronic inhomogeneities in the superconducting dome or even the possible microscopic coexistence of magnetic ordering and superconductivity in the FeAs layers,\cite{Amato} which might have eluded previous neutron diffraction studies.\cite{Zhao2008}
In fact, contrary to diffraction techniques, which cannot detect short range magnetic order, muons act as local magnetic probes, hence making muon spectroscopy ($\mu$SR) an ideal tool for this sort of investigations.
For this reason $\mu$SR has long been employed to study the M-SC coexistence in cuprates \cite{Savici,Sanna2004,Russo,Keren2003,Coneri2010} as well as in other superconducting compounds, such as the ruthenocuprates,\cite{Bernhard1999} or the heavy-fermion superconductors.\cite{Llobet,Park,Visser}

Here we report on zero- (ZF) and transverse-field (TF) $\mu$SR measurements on a sample of \cefaof\, which unambiguously show the coexistence of superconductivity and short range magnetic order on a nanoscopic length scale. While in contradiction with previous experimental findings on the same compound, \cite{Zhao2008} this result closely resembles the behavior of Sm1111 at the M-SC crossover.\cite{Drew2009,Sanna2009,Hess2009,Kamihara2010}

The investigated polycrystalline \cefaof\ sample was synthesized by a solid-state reaction method
%using CeAs, Fe, CeO$_2$ , CeF$_3$ , Fe$_2$As as starting materials. CeAs was obtained by reacting Ce chips and As pieces at 500 C for 15 h and then 850 C for 5 h. The raw materials were thoroughly mixed andpressed into pellets. The pellets were then annealed at 1150 C for 50 h. The resulting samples were characterized by a powder x-ray diffraction (XRD). Further details concerning sample preparation and characterization have been
following the procedure reported in Ref.~\onlinecite{Palenzona2010}. The total fluorine content was evaluated from intensity measurements of the $^{19}$F nuclear magnetic resonance echo signal, as compared to that of a SmOF reference compound.
Successive Rietveld analysis of the powder x-ray diffraction pattern excluded the presence of fluorine in other secondary phases, except for a tiny minority (3\% vol.) of a spurious CeOF phase.
The combined result of the above analysis gives a best estimate of $x=0.06(2)$ for the F stoichiometry in \cefaof.

\begin{figure}
\includegraphics[width=0.4\textwidth,angle=0]{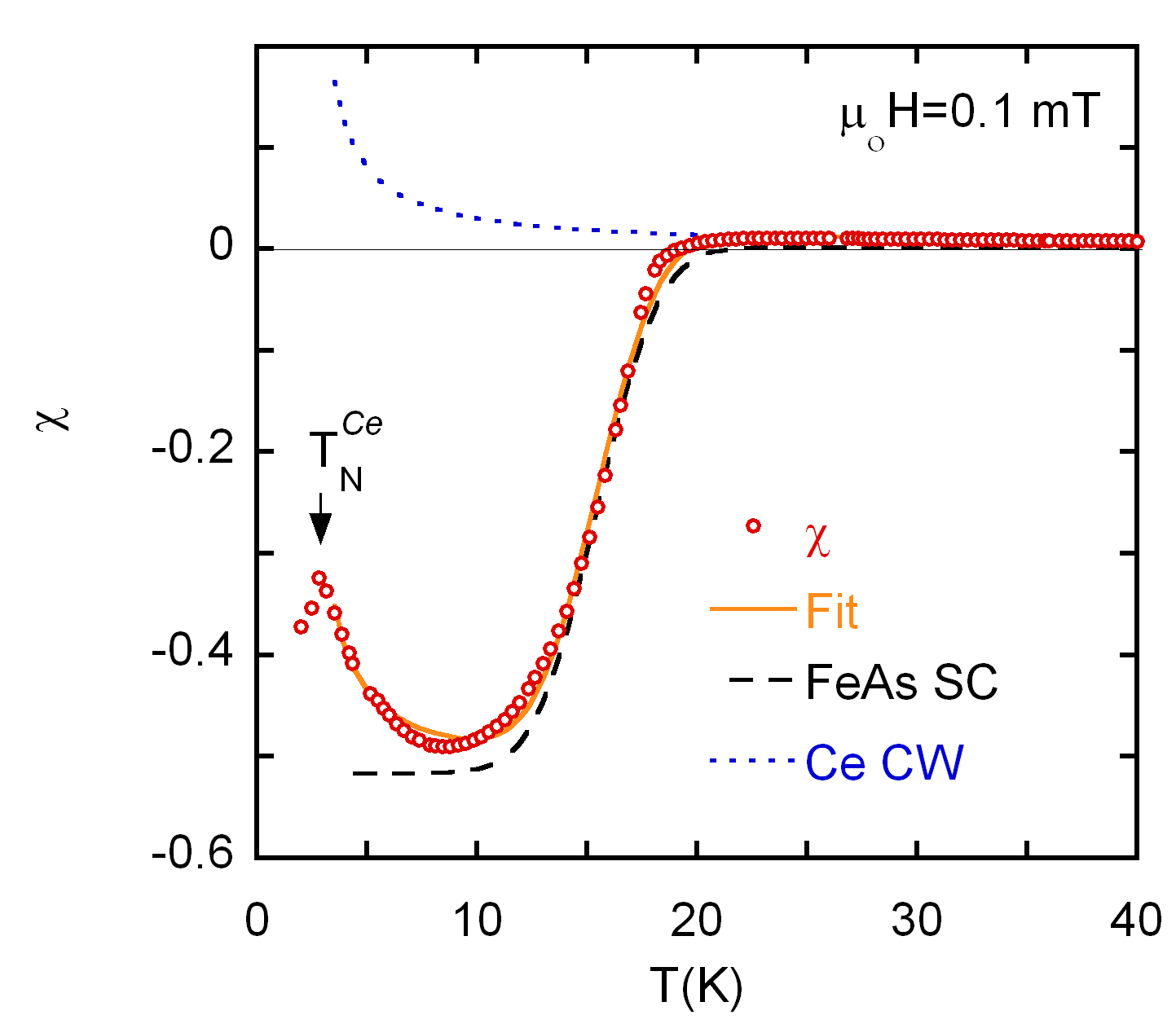}%
%[trim=  15 170 15 170, clip,
\caption{\label{fig:squid}(Color online)  Magnetic susceptibility in zero-field cooling of
\cefaof\ with $x=0.06(2)$, plotted in SI units. The solid curve represents a phenomenological fit to the susceptibility for temperatures above the ordering of Ce magnetic sublattice $T_N^{\mathrm{Ce}}$. The superconducting (dashed line) and the Curie-Weiss (dotted line) contributions are also displayed (see text for details).}
\end{figure}

The temperature dependence of magnetic susceptibility, $\chi(T)$, was measured on the powder sample
using a dc Superconducting Quantum Interference Device (SQUID), and it is shown in Fig.~\ref{fig:squid}. Two key features are
evident from the data: a sizeable diamagnetic response below $T_c=18$ K due to SC shielding, and a cusp at $T_N^{\mathrm{Ce}}=2.9$ K due to the antiferromagnetic ordering of the Ce sublattice.\cite{Chen2008,Maeter2009} A similar behavior is found in an optimally doped Ce1111 sample.\cite{Chong2008}
To empirically separate the contributions due to the electrons in FeAs bands from the ones of Ce$^{3+}$, the susceptibility was fitted to the sum of two functions:
an $\mathrm{erf}[(T-T_c)/(\sqrt{2}\Delta)]$, which accounts for the superconducting transition (at $T_c$ with a width $\Delta$), and a Curie-Weiss term, which accounts for the behavior of the Ce sublattice.
% risultati del fit: c=0.23, theta=2.1 K; a=0.259; b=.993; Tc=15.4 K; Delta=2.2 K
The two contributions are shown in Fig.~\ref{fig:squid} by dashed and dotted lines, respectively.
From the low-temperature limit of the first term, $\chi_{\mathrm{sc}}(T\rightarrow0)\simeq 0.5$ (in SI units),
one can roughly estimate a $\sim 50\%$ superconducting volume fraction.

This fraction could be even larger, since at low doping the field penetration depth increases considerably,\cite{Carlo2009} and becomes comparable to the grain size (1--10 $\mu$m). Hence the shielding volume is effectively reduced within each grain. The SC fraction could also be smaller if superconductivity were limited to the grain surface, but we shall show this not to be the case by TF-$\mu$SR.
%This superconducting behavior is very similar to that observed in Sm1111, which (apart from an analogous low-$T$ ordering of Sm ions) displays also magnetic ordering in the FeAs layers.\cite{Sanna2009}

\begin{figure}
\includegraphics[width=0.4\textwidth,angle=0]{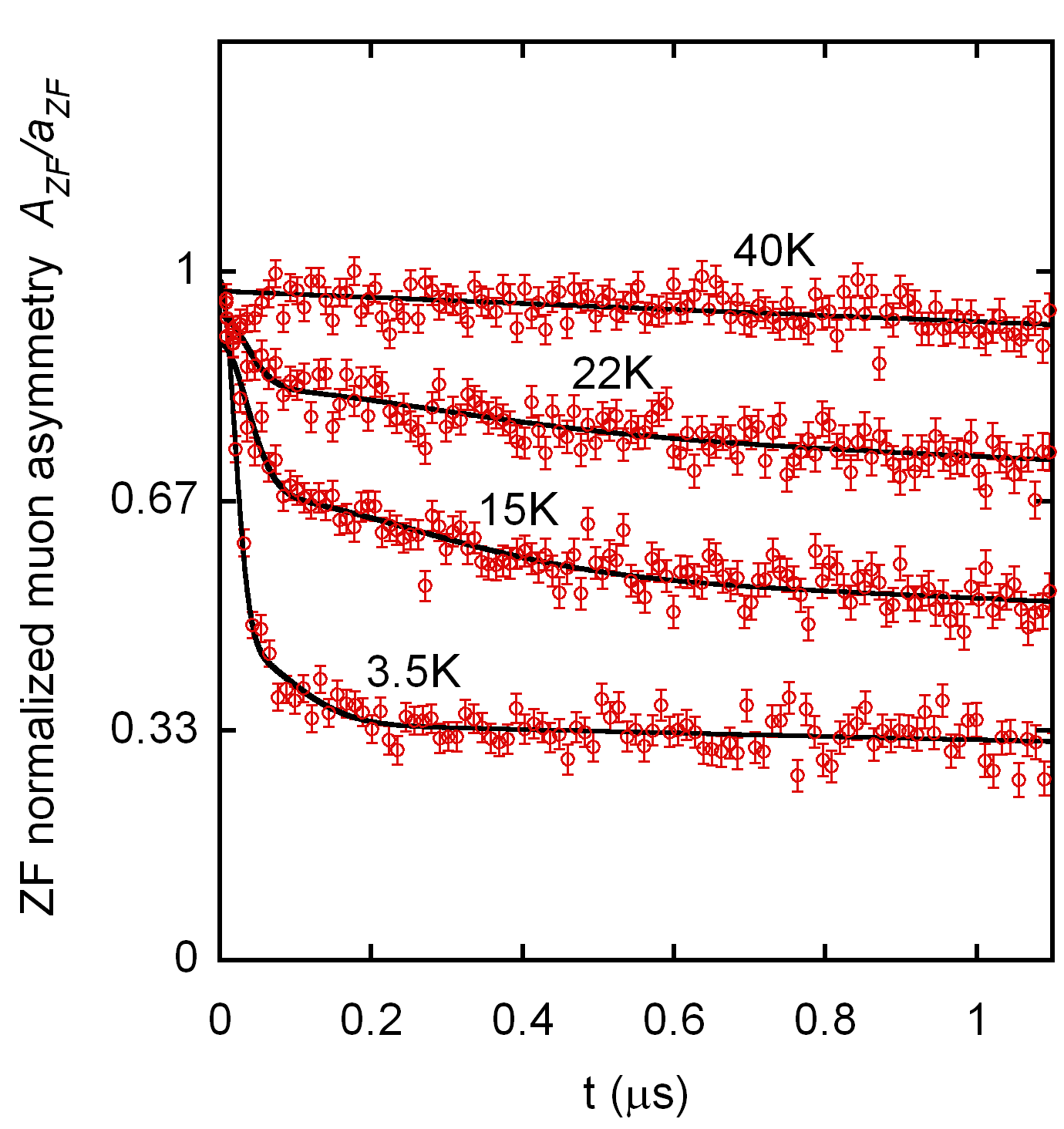}%
\caption{\label{fig:asymmetry}(Color online) Time dependence of the normalized zero-field muon asymmetry with best fits to Eq.~(\ref{eq:ZFasymmetry}), measured at four different temperatures.}
\end{figure}

To probe the local magnetic state in Ce1111 we performed a series of ZF-$\mu$SR measurements. Figure~\ref{fig:asymmetry} shows the time dependence of the ZF muon asymmetry, ${\cal A}_{\mathrm{ZF}}(t)$, normalized to its room temperature value $a_{\mathrm{ZF}}$
(a marginal muon fraction of 5\%, due either to muons stopped in the cryostat walls or in a non magnetic impurity phase, was subtracted as a constant background). Solid lines show the best fit to the measured sample asymmetry using the following normalized ZF function:
\begin{equation}
\frac {{\cal A}_{\mathrm{ZF}}(t)} {a_{\mathrm{ZF}}} = f_L \, e^{-\lambda_L t} + f_T \cdot ( w_1 e^{-\sigma^2_1 t^2/2} + w_2 e^{-\sigma^2_2 t^2/2})
\label{eq:ZFasymmetry}
\end{equation}
Here we distinguish a slowly decaying ($\lambda_L \sim 0.06$ $\mu$s$^{-1}$) muon fraction, $f_L$, whose amplitude increases from 1/3 at low temperature to a unitary value at high $T$, and a second muon fraction, $f_T$, which vanishes at high temperature. One can easily identify them with the longitudinal ($\bm{B}_i\!\parallel\! \bm{S}_\mu$) and transverse ($\bm{B}_i \!\perp\! \bm{S}_\mu$) components of the asymmetry, respectively, with $\bm{B}_i$ the internal magnetic field and $\bm{S}_\mu$ the initial muon-spin direction.

\begin{figure}
\includegraphics[width=0.45\textwidth,angle=0]{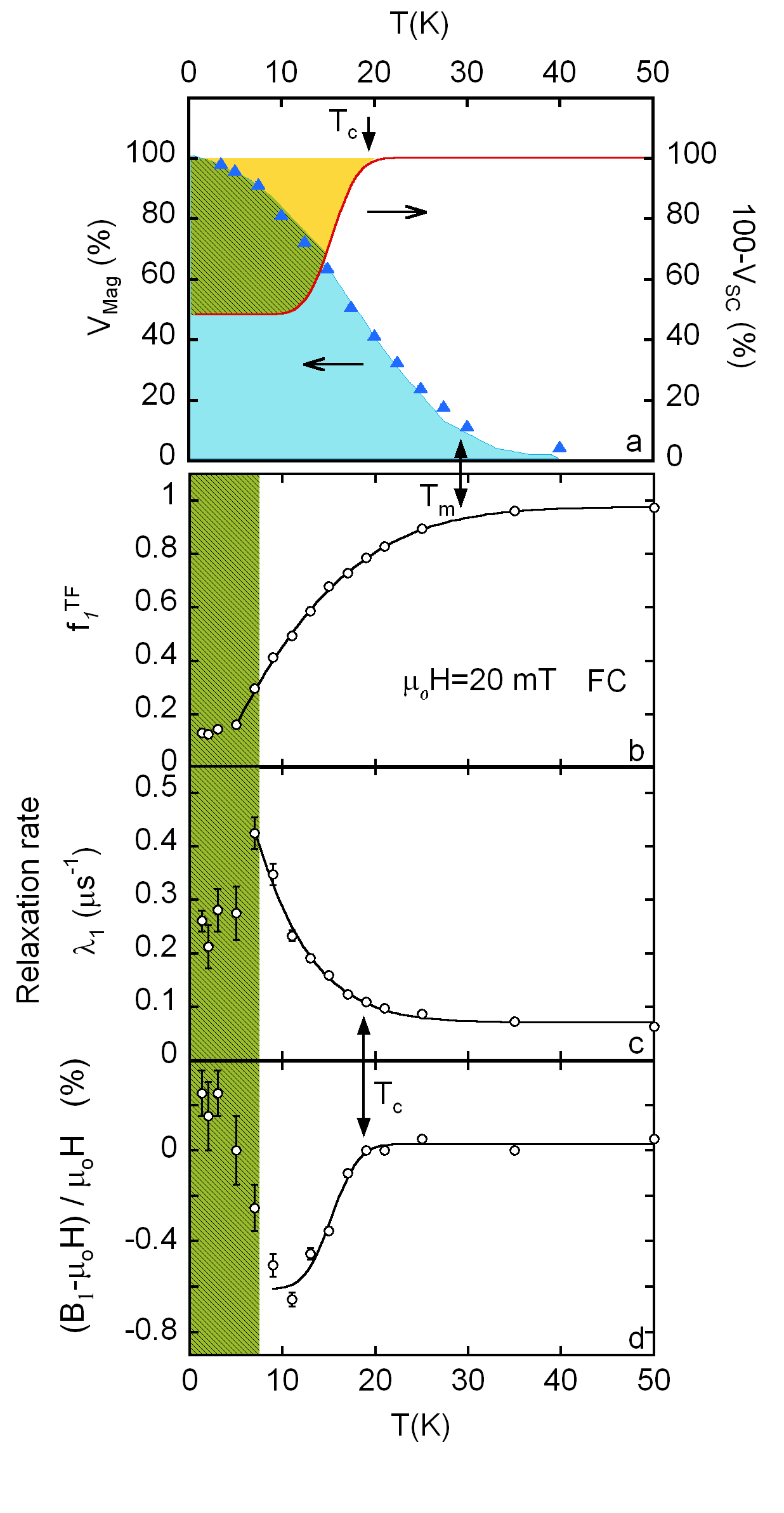}
\caption{\label{fig:fig3e4}(Color online) a) Temperature dependence of the magnetic (triangles) and non-superconducting (solid line) volume fractions as seen by ZF-\-$\mu$SR and magnetization measurements, respectively. The onset of superconducting and magnetic transitions, $T_c$ and $T_m$, is indicated by vertical arrows. b), c) and d) panels display the fraction, the decay rate and the relative field shift for the $j=1$ fraction of the TF-\-$\mu$SR asymmetry (Eq.~\ref{eq:Tasymmetry}).}
\end{figure}

The very fast relaxing transverse components represent the signature of a sizeable distribution of internal fields $\bm{B_i}$.  Best fits at low temperature yield two Gaussian contributions with weights $w_1=0.85$ and $w_2=0.15$ and standard deviations $\sigma/2\pi\gamma\!\!=\!\!(\overline{B_i}^2- \overline{B_i^2})^{1/2}\simeq60$ mT and 12 mT, respectively. Internal fields of this size are typically found at the muon site when the magnetic ordering occurs in the FeAs layers of samples close to a M-SC crossover.\cite{Sanna2009} Indeed, since we find both the transverse components to disappear at the same temperature, they should reflect the same electronic environment. These two transverse components most probably come from two different muon stopping sites as suggested by a previous $\mu$SR study in undoped Ce1111 samples.\cite{Maeter2009}
By considering that simple geometric arguments predict $f_L=1/3$ for a fully AF ordered polycrystalline sample, we can estimate
the magnetic volume fraction due to FeAs layers as $V_{\mathrm{mag}}=3(1-f_L)/2$.

The temperature dependence of $V_{\mathrm{mag}}$ is reported in Fig.~\ref{fig:fig3e4}a.
It shows that the magnetic transition has its onset already at $T_m \simeq 30$ K and that the whole sample
becomes magnetic below $T \lesssim 10$ K, hence proving the presence of ordered magnetic moments throughout the FeAs layers of the {\em whole} sample volume. This does not necessarily imply that all the muons are implanted inside a magnetically ordered domain.
The distance between adjacent antiferromagnetic domains (i.e. with vanishing macroscopic moment) can be estimated by simply considering the dipolar interaction between the $S_\mu = \frac{1}{2}$ muon spin and a domain moment with the value of the ordered moment, $m \approx 0.3 \mu_\mathrm{B}$,\cite{Klauss2008} which at a distance $d$ produces a local field $B_i=\frac{\mu_0}{4\pi} md^{-3}$.
Since in ZF-$\mu$SR a rough detection limit for the spontaneous internal fields
is ca.\ 1 mT, one can estimate to $d\sim$1 nm the maximum ``detectable'' distance between an ordered domain and a muon site. Considering now that in our Ce1111 sample practically every muon experiences a non vanishing local field from the FeAs layer for $T\lesssim10$ K (see Fig.~\ref{fig:fig3e4}a), one can conclude that the maximum distance between magnetically ordered domains is of the order of a few nanometers.
Combined with the above SQUID measurements, the ZF $\mu$SR results clearly demonstrate that at low temperature the SC and M states coexist within a {\em nanoscopic} length scale in at least 50\% of the sample volume, as shown by the hatched area of Fig.~\ref{fig:fig3e4}a.
This coexistence implies that the superconductivity must survive within a few nanometers, a condition which is satisfied in this material, where the typical coherence length is of the order of $\xi \sim 2$ nm.\cite{Chong2008}

To further investigate the M-SC coexistence state we carried out TF-$\mu$SR measurements, whereby the sample was cooled in an externally applied field $\bm{H}\perp \bm{S}_\mu$ equal to $\mu_0 H=20$ mT, i.e.\ higher than the lower superconducting critical field $H_{c1}$, expected in the range 0--10 mT.\cite{Okazaki2009} Accordingly, a flux-line lattice is generated below $T_c$. In this experiment muons probing the {\em pure} flux-line lattice experience the diamagnetic shift of the local field $B_\mu\!\!=\!\! \mu_0 H(1+\chi)$, with $\chi<0$.\cite{Brandt1988} On the other hand, those muons implanted in the magnetically ordered phase will probe a magnetic field $B_\mu\!\!=\!\! |\mu_0\bm{H} + \bm{B}_i|$, whose magnitude in a powder sample is $B_\mu \!\gtrsim\! \mu_0 H$.\cite{Allodi2006} The amplitudes of these %two
frequency-distinct signals are proportional to the volume fractions where the corresponding order parameter is established. Based on these considerations, we could describe the time evolution of the TF-$\mu$SR normalized asymmetry using:
\begin{equation}
\label{eq:Tasymmetry}
\frac {{\cal A}_{\mathrm{TF}}(t)} {a_{\mathrm{TF}}} \!=\! \sum_{j=1,2} f^\mathrm{TF}_j e^{-\lambda_j t}\cos(2\pi\gamma B_j t) + f^\mathrm{TF}_{3} e^{-\lambda_3 t},
\end{equation}
with $\gamma\!=\!135.5$ MHz/T, the muon gyromagnetic ratio and $a_{\mathrm{TF}}$, the total asymmetry measured at high temperature.

Equation~(\ref{eq:Tasymmetry}) fits the TF data very well over the entire 3--300 K temperature range ($\chi^2\approx1\div1.2$). The last non-oscillating term accounts for the longitudinal component of the muon spin in the magnetically ordered phase $(\mu_0\bm{H} + \bm{B}_i) \!\parallel\! \bm{S}_\mu$, expected below $T_m$. %Equation~(\ref{eq:Tasymmetry}) gives very good fits over the entire temperature range ($\chi^2\approx1\div1.2$).
The second of the oscillating terms (the one labeled with $j=2$ --- not shown), is present only below $T_m$. It reflects an environment with spontaneous magnetic order, characterized by paramagnetic field shifts at the muon site $B_2 \approx 23$ mT ($>\mu_0H$), and by fast ($\lambda_2 \sim 5$ $\mu$s$^{-1}$) relaxation rates due to the disordered distribution of spontaneous
local fields $B_i$, in agreement with previous ZF-$\mu$SR results.

Let us now focus on the parameters describing the first ($j=1$) oscillating term. Figure~\ref{fig:fig3e4}b shows the
fraction $f^\mathrm{TF}_1$ that  is close to one at high temperatures (with $f^\mathrm{TF}_{2}=f^\mathrm{TF}_{3}=0$), since the whole sample is in a single phase for $T>T_m$.
Interesting insights come from the relative field shift sensed by implanted muons (shown in Fig.~\ref{fig:fig3e4}d).
In this high-$T$ regime the absence of a shift characterizes a sample which is neither in a superconducting, nor in a magnetically ordered state. Here the Lorentzian character of relaxation, with small $\lambda_1 \lesssim 0.1$~$\mu$s$^{-1}$ values (see Fig.~\ref{fig:fig3e4}c), reflects the presence of very small fluctuating dipolar fields, probably due either to the Ce magnetic moments or to some minor phase of diluted Fe clusters.\cite{SannaJSC2008} Once the sample is cooled below $T_m$ a reduction of $f^{\mathrm{TF}}_1$ is observed, specular to the increase in magnetic volume fraction detected by ZF-$\mu$SR, as clearly seen in panels a and b of Fig.~\ref{fig:fig3e4}.
However, no appreciable variations in $\lambda_1$ or $B_1$ are detected across $T_m$, suggesting that no electronic changes occur in the $f^\mathrm{TF}_1$ volume fraction down to $T_c$.
Only below $T_c$ there is a sizeable increase of the diamagnetic shift (panel d), which denotes an expulsion of the externally applied field, as well as the increase of the relaxation rate (panel c), which reaches values typical of the superconducting pnictides.\cite{Carlo2009} Notice that the muon fraction in the superconducting environment is $f^\mathrm{TF}_1>0.5$ for $10\,\mathrm{K}< T < T_c$, which demonstrates that the corresponding volume is more than 50\%. By further cooling below 10 K (hatched area in panels b--d) one finds that  $f^\mathrm{TF}_1$ reduces drastically to $\sim 15\%$. Interestingly, there is also a simultaneous drop in the relaxation rate $\lambda_1$ and a progressive vanishing of the diamagnetic shift $B_1$. All these facts imply that the magnetic environment probed by muons is far more complex than the initially pure flux-line lattice, with internal fields $B_i$ of the order of $\mu_0H$ developing throughout the whole volume within a nanoscopic length scale.
This picture fully agrees with that from ZF-$\mu$SR, also consistent with the presence of coexisting magnetic order in the FeAs layers.

In summary, both ZF- and TF-$\mu$SR experiments show that a superconducting Ce1111 sample becomes fully magnetic within the FeAs layers below 10 K. Below $T_c$ a sizeable fraction of muons detect a pure superconducting volume, which seems to progressively vanish as the fully ordered magnetic state develops. This, however, does not imply that superconductivity is destroyed, as clearly proved by susceptibility measurements, which detect a practically unchanged SC volume fraction (once the unrelated paramagnetic behavior of Ce is properly accounted for).

These results demonstrate that in Ce1111 the superconductivity may coexist at the nanoscopic scale with magnetically ordered moments in the FeAs layers. This means that the magnetic and superconducting order parameters cannot vanish simultaneously, in contrast with earlier studies, \cite{Zhao2008} hence excluding the presence of a common quantum critical point.\cite{Amato} Indeed this behavior closely resembles that of Sm1111,\cite{Sanna2009} suggesting that the coexistence of magnetism with superconductivity within the FeAs layers is a feature common to different RE1111 pnictides. Further studies are necessary to measure the extent of the region of M-SC coexistence in Ce1111 as a function of F doping.

%\begin{acknowledgments}
\textbf{Acknowledgments} $\mu$SR experiments were carried out at the PSI and ISIS muon facilities. We acknowledge partial financial support from the PRIN-08 project and from PSI EU funding.
%\end{acknowledgments}

%\bibliography{Ce1111}

\end{document}